\documentclass[letter,fleqn,usenatbib]{mnras}

\usepackage{newtxtext,newtxmath}

\usepackage[T1]{fontenc}
\usepackage{ae,aecompl}

\usepackage{graphicx}	
\usepackage{amsmath}	
\usepackage{amssymb}	

\title[Primordial Segregation of Mass and Density]{Primordial Mass and Density Segregation in a Young Molecular Cloud}

\author[Alfaro \& Rom\'an-Z\'u\~niga]{
Emilio  J. Alfaro,$^{1}$\thanks{E-mail: emilio@iaa.es}
and Carlos G. Rom\'an-Z\'u\~niga,$^{2}$
\\
$^{1}$Instituto de Astrof\'isica de Andaluc\'ia, (CSIC), Glorieta de la Astronom\'ia, S/N, Granada, 18008, Spain\\
$^{2}$Instituto de Astronom\'ia, UNAM, Unidad Acad\'emica en Ensenada, Ensenada 22860, Mexico\\
}

\date{Submitted for publication in MNRAS, March 2018}

\pubyear{2018}

\begin{document}
\label{firstpage}
\pagerange{\pageref{firstpage}--\pageref{lastpage}}
\maketitle
%
%
%
%
\begin{abstract}
We analyse the geometry of the Pipe Nebula, drawn by the distribution ($Q$-spatial parameter) and hierarchy ($\Lambda$ spatial segregation) of column density peaks previously detected and catalogued. By analysing the mass and volume density of the cores, we determine that both variables shown to be spatially segregated and with a high degree of substructure. Given the early evolutionary state of the Pipe Nebula, our results suggest that both, mass and volume-density segregations, may be primordial, in the sense of appearing in the early phases of the chain of physical mechanisms which conform the cluster-formation process. We also propose that volume density, and not mass, is the parameter that most clearly determines the initial spatial distribution of pre-stellar cores. 
\end{abstract}



\section{Introduction}\label{intro}

\par Most star clusters in the Galaxy have their most massive stars concentrated near their centres. This spatial segregation of the mass has been thought to be  a consequence of an equipartition of energy driven by gravitational stellar encounters \citep[e.g.][]{meylan:2000,khalisi:2006}. This is a well-known dynamic process whose evolutionary time depends on the average mass of the cluster stars, on the mass of the spatially segregated stars, and on the relaxation time of the system. The mass segregation effect in stellar clusters does not appear to depend of local environment \citep[e.g.][]{dib:2018} or the initial dynamical configuration \citep[e.g.][]{dominguez:2017}. However, many young clusters with relaxation times longer than their age also show evidence of being mass-segregated, and in those cases, the spatial pattern of the mass cannot be explained by the standard two-body relaxation mechanism \citep{Parker+16}.

\par The formation of stellar clusters is far from being a well-understood process. On the one hand we have the progenitor giant molecular clouds whose initial conditions in terms of structure in density, temperature and velocity seem to be well known and limited for the most part to a narrow range of values \citep[e.g. ][]{sanchez:2005aa, sanchez:2007aa, andre:2010aa, andre:2013aa}, and, on the other hand, we have the recently formed stars that are also constrained observationally by an Initial Mass Function (IMF) with a probable universal character \citep{padoan:1997aa, bastian:2010aa, ascenso:2012aa, parker:2012aa, PSA12}, but whose distribution in phase space shows a more variable casuistry \citep{sanchez:2009aa, elmegreen:2009aa}. In between these two states there are several phases: a) the magneto-hydrodynamic evolution of the cloud towards the formation of pre-stellar nuclei; b) the process of perturbation, collapse and fragmentation of the pre-stellar nuclei that gives place to proto-stellar nuclei; c) the accretion and/or coalescence of the proto-stellar nuclei favoured or restricted by the density of objects, the magnetic field and turbulence and finally d) a group of stars with a characteristic mass distribution undergoing a rapid dynamic evolution in a medium not always free of interstellar matter \citep{allison:2010aa, parker:2013aa, bressert:2010aa}. Additional complications arise from two facts: the progression of cluster formation and early evolution is intimately linked to the primordial structure of the forming cloud \citep[e.g.][]{roman:2008aa,ybarra:2013aa}, and that clusters do not form as isolated entities, but they form in families across clouds, and present abundant substructure \citep[e.g.][]{roman:2015aa, GA17}. 

\par One question we may able to answer is: When and how the observed mass segregation in young clusters is then generated? In this paper we study the spatial structure of a list of pre-stellar column density peaks identified across the Pipe Nebula \citep{roman:2010aa}. This object is a very young cloud, apparently dominated by a long quiescent state \citep{lada:2008aa}, with only one core actively forming a small, low-mass stellar cluster \citep{brooke:2007aa} and less than a handful of additional well defined cores showing infrared emission indicative of proto-stellar activity \citep{forbrich:2009aa}. Thus, the Pipe Nebula probably represents an early evolutionary state of the star-formation process. The cloud is possibly at a stage when its structure and evolution is controlled by magneto-hydrodynamic physics but just before the onset of the cloud collapse when gravitation is becoming the main driving force. The Pipe Nebula thus sets a lower evolutionary limit for the study of the presence or absence of mass segregation in the earliest stages of the formation of a stellar cluster.  The column density peak catalogues of \cite{roman:2009aa,roman:2010aa} (hereafter RALL0910) are particularly useful for this study. The catalogues were constructed from a set of high resolution column density maps with a large dynamic range ($N_{H_2}<8.4\times 10^{22}\mathrm{\ cm}^{-2}$) that allowed to detect significant structures down and below the Jeans length, resulting in a very complete census of the pre-stellar component of the cloud. 

\par This paper is divided into four sections, the first being this introduction. Section \ref{catalog} is devoted to the description of the Pipe Nebula and to the analysis of the physical properties of the pre-stellar cores forming our observational sample. In Section \ref{method} we analyse the spatial pattern of the cores in function of different physical properties, and, finally (Section \ref{results}), we discuss our results in the context of the formation of star clusters.

\section{Column density peak catalogue}\label{catalog} 

\par
\begin{figure}
\includegraphics[width=0.45\textwidth, angle=0]{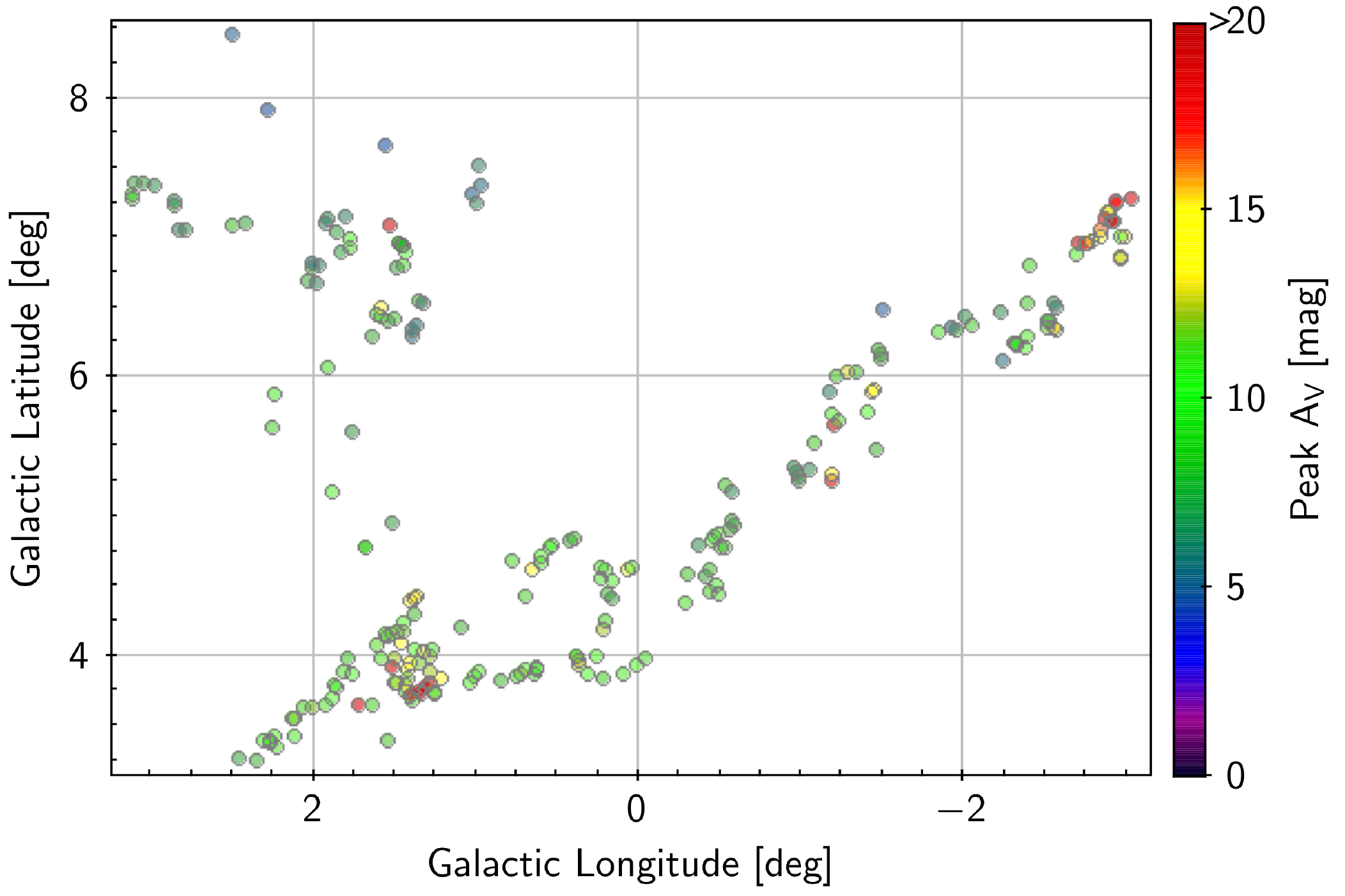}
\caption{A map of column density peak positions in the Pipe Nebula. The color scale indicates the peak value of visual extinction, $A_V$, in magnitudes. At first glance, higher extinction peaks appear located in the most populated regions of the map.}
\label{fig:map}
\end{figure}

 In this study we make use of column density peak catalogues of RALL0910 obtained from near-infrared, high-resolution\footnote{The maps were constructed with a Gaussian beam of 20$\arcsec$ FWHM} visual extinction maps of the Pipe Nebula. These maps were constructed using the Near-Infrared Color Excess Method \citep{lombardi:2006aa} applied to a combination of various near-infrared ($J$,$H$ and $K_s$ bands) photometry catalogues from surveys performed at ESO-NTT, ESO-VTL, and Calar-Alto-3.5m telescope, complemented with photometry from the Two-Mass All-Sky survey (2MASS). For a fully detailed description of the maps, we refer the reader to RALL0910. Summarising the methods, significant column density peaks were detected and classified trough the application of the \texttt{CLUMPFIND} algorithm \citep{williams:1994aa} to background subtracted versions of the column density maps. Background subtraction was applied by filtering large scale emission using a wavelet based technique \citep{rue:1997}. The catalogue of RALL0910 contains a list of 244 column density peaks. For each peak they provided: a) the position of the peak, defined at the pixel with the highest value, b) the peak mass, obtained from a conversion from extinction to total mass of H$_2$ using a constant gas to dust ratio, and a distance of 130 pc to the cloud, c) an equivalent radius (obtained from the area of each region, defined by \texttt{CLUMPFIND}), and d) a mean volume density, estimated from the radius and mass assuming a simple spherical geometry. In Figure \ref{fig:map} we show the spatial distribution of the column density peaks across the cloud. 

\par It is important to clarify here that not all of these column density peaks may be considered as individual pre-stellar cores. The main reason (also discussed by RALL0910) is that we do not have gas kinematics information (either radial velocities or velocity dispersions) for all individual peak positions in order to determine if all objects in our list are independent. As prescribed by \citet{rathborne:2009aa}, in order to determine if a column density peak is an independent core, we would require to comply that the separation from a neighbour peak is equal or larger than the Jeans length (for instance, at $T=10$ K and $n=10^3\mathrm{\ cm}^{-3}$) and that the radial velocity difference from peak to peak is equal or larger than the local sound speed. Thus, as mentioned by RALL0910, a significant fraction of our peaks appear to be genuine sub-Jeans structures but we cannot fully identify them as independent pre-stellar cores. However, nowadays we have confirmed that significant peaks can be found with sub-Jeans sizes and separations in column density maps obtained from Herschel\footnote{Herschel is an ESA space observatory with science instruments provided by European-led Principal Investigator consortia and with important participation from NASA.} far-infrared emission observations. Sub-Jeans sized or sub-Jeans structured peaks may play a role in the early evolution of the molecular gas of a cloud like the Pipe Nebula: current discussions suggest that proto-stellar regions in star-forming molecular clouds may be the result of the merging of subsidiary material that flows along filaments \citep[e.g][]{LHC17,roca:2015,gwilliams:2018}. In this sense, sub-Jeans peaks may be able to merge into classical Jeans-sized pre-stellar cores, or even to survive as independent low-mass globules. For these reasons, we consider that it is important to use the complete list of column-density peaks for our analysis, as we are precisely, looking for a significance of the spatial structure driven by different physical variables, along the whole extension and along the various scales of the cloud.

\subsection{Spatial distribution of column density peaks}

\par
\begin{figure}
\includegraphics[width=0.42\textwidth, angle=0]{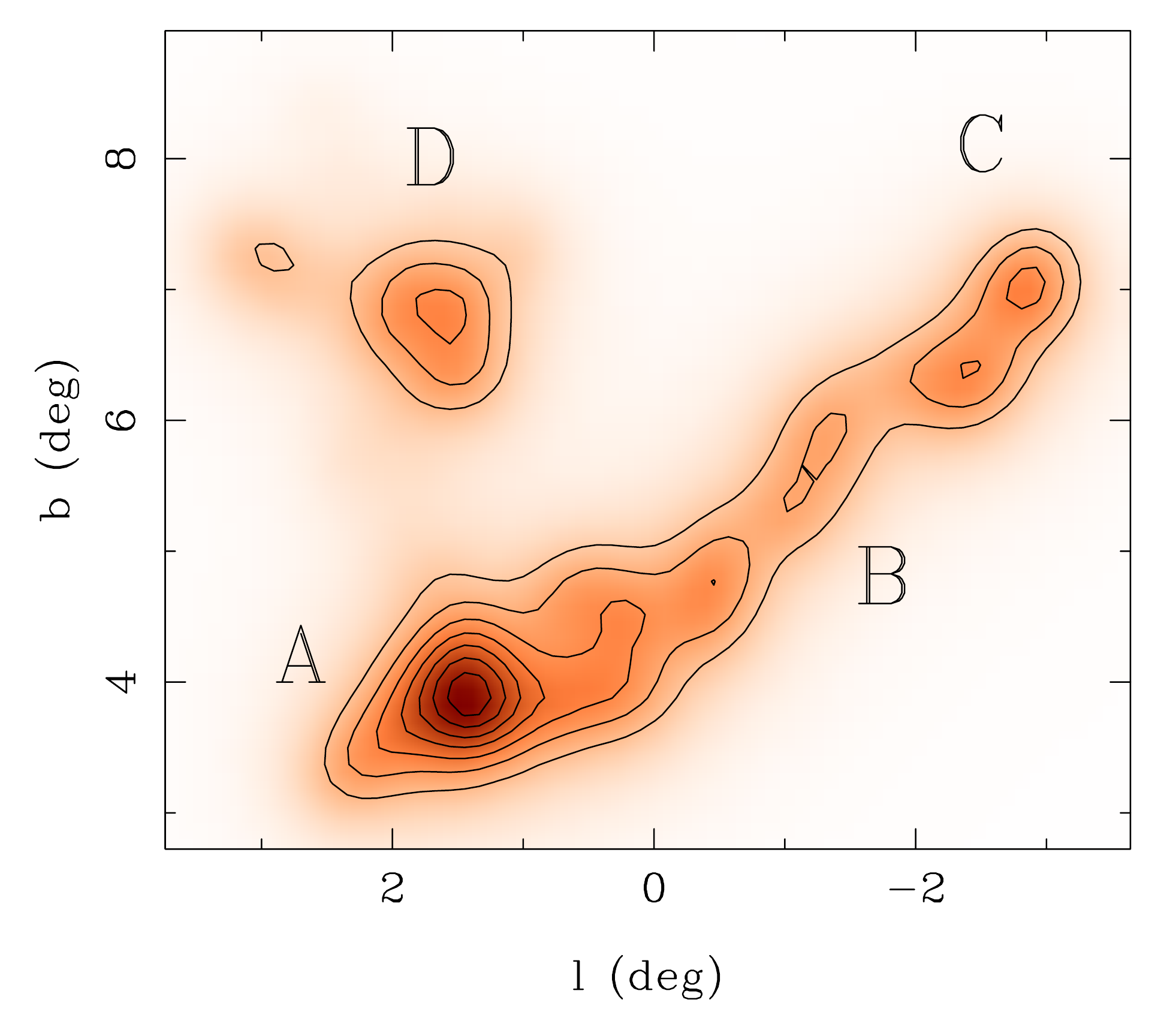}
\caption{Surface density map or Pipe nebula peaks. The labels indicate four main concentrations of peaks across the cloud. Notice the three higher concentrations are located at the tips of the filaments.}
\label{fig:surfdens}
\end{figure}

 In Figure \ref{fig:surfdens}, we present a ($l, b$) surface density map of the column density peaks in our sample. We can distinguish four well-defined concentrations (labeled as A, B, C and D), the largest and most dense (A) being found at the intersection of the two lineal filaments drawn by the cores distribution. This region is known as \textit{The Bowl}, has been shown to be consistent with the collisional merging of two well defined molecular filaments permeated by a magnetic field \citep{frau:2015aa}. It contains a high concentration of peaks, some of them with peak values above $A_{V} > 20$ mag \citep{roman:2009aa}. Two other concentrations appear aligned along the denser and longer filament: the one labeled as B in Figure \ref{fig:surfdens} is known as \textit{The Stem}, that shows the most obvious filamentary morphology in the cloud. The core concentration defining the western end of the filament contains the clump Barnard 59 (label C in the figure), is the only region in the cloud with active star formation. About twenty proto-stars are embedded in a region of very high column density, reaching in its centre values of $A_V$ for above several tens of magnitudes. Finally, the concentration labeled as D in the figure coincides mostly with \textit{The Smoke} region of the Pipe, where the well known core Barnard 68 \citep{alves:2001} is located.

\par We notice that this figure is equivalent to Figure 13 discussed in section 5.2.1 of \citet{roman:2010aa}, except that the current map is made with a lower resolution grid in order to enhance the four main concentrations discussed above. The fractal dimension of this distribution has been estimated to be $1.23\pm 0.02$, very representative of the filamentary morphology of the cloud. A similar map by In the paper by \citeauthor{roman:2010aa}  showed a very well defined tendency for extinction peak grouping, suggestive of a possible primordial origin for clustering.

\subsection {Peak mass and density distribution}

\par Concerning other core properties analysed in this study, the Core Mass Function (CMF) of Pipe's A$_{V}$ peaks has been fitted by a Chabrier differential mass  function \citep{chabrier:2005}, defined as a lognormal distribution for the lower mass cores, while the higher mass tail is well represented by a Salpeter power law \citep{salpeter:1955}. Using our sample, we obtained a slope of 1.38$\pm$0.11 for the log-log Salpeter's function and a lognormal distribution with mean $\mu (\log{(\mathrm{M\ [M_\odot]})})=-0.38$ and dispersion $\sigma ( {\log{(\mathrm{M\ [M_\odot]})}})=0.45$. The separation between both regimes was chosen at M$_{limit}$ = 0.6 M$_\odot$. This is compatible with the CMF determinations of \citet{alves:2007,rathborne:2009aa} and confirms that the Pipe behaves as most nearest clouds do, to say, as the initial mass of stellar cluster members distributes, but shifted in mass by a factor of 3 \citep{alves:2007, nutter:2007, enoch:2008}. If we compare our CMF with other nearby cloud as Aquila \citep{konyves:2010} we find that the mean and dispersion of the lognormal part of the differential mass function for Aquila starless cores, (-0.22, 0.42), are very similar to the values obtained in this analysis. The starless cores in the Aquila cloud are  more massive than the peaks in the Pipe Nebula by about 0.2 M$_{\odot}$, in average.  Salpeter's  exponent for Aquila is 1.50$\pm$0.2, compatible with our estimate for the Pipe.  The core volume density histogram for our sample of density peaks in the Pipe are roughly represented by a lognormal function with mean $\mu (\log{(\rho\mathrm{[cm^{-3}]})})=4.58$,  $\sigma ( {\log{(\rho\mathrm{[cm^{-3}]})}})=0.19$. The central value is very similar to the density distribution for starless cores in the Taurus molecular cloud \citep{marsh:2016}. Thus, for both mass and volume density distributions for the Pipe A$_V$ are very similar to those of other molecular clouds in the solar neighbourhood analysed with {\sl Herschel} space mission data. 

\section{Method for spatial segregation analysis}\label{method}

\par Several algorithms for analysing mass segregation in stellar systems have been proposed, being particularly useful those that apply a direct measurement of radial variations of surface density or the mass distribution itself \citep[e.g. ][]{degrijs:2002,gouli:2004}. Another frequently used method relies on the properties of the Minimum Spanning Tree (MST) edge length distribution. This approach to the mass segregation analysis was formerly introduced by \cite{allison:2010aa} and it is founded on the idea that we find mass segregation if massive stars are closer to other massive stars than to any other star. This definition does not need of a spatial singular centre around which the massive stars have to be concentrated, but they can form clustering around any non-singular position. 

\par We used the MST method on the Pipe Nebula density peak list. Our procedure can be easily summarised as follows: a) Cores are sorted according to their masses in descending order; b) An interval of size $n$ is defined as the number of objects per bin; c) The MST of the spatial distribution for the $n$ cores in the interval is calculated, from which the central value of the edge length distribution, $l$, is obtained; d) this value $l$ is compared to a value $l_r$ obtained for a random sample of the same size ($n$) taken from the complete sample. In order to get a representative value of this central estimator, $l_r$, we obtained the average of multiple repetitions; e) We take the quotient between $l$ and $l_r$ ($\Lambda$, following the notation of \citeauthor{allison:2010aa}) for the case where $l$ is the central length. If $\Lambda>1$ we say we are observing mass segregation. We also select a step parameter $s$, which establish the number of shifted objects between two consecutive bins

\par In the same way, we estimated the $Q$-spatial parameter \citep{C&W2004} for each bin, which quantifies the projected spatial structure of $n$ objects per bin, sorted either by mass or density (hereinafter $M$ and $\rho$, respectively). $Q\lesssim0.8$ values are indicative of a clumpy, sub-clustered structure, whereas $Q\gtrsim0.8$ indicate a more centrally condensed distributions. The further the value of Q from 0.8, the stronger the degree of sub-clustering or central concentration. There exist clear relationships between $Q$ and fractal dimension, and $Q$ and King's concentration parameter for radial distributions \citep{sanchez:2009aa}. It is important to notice that if we were facing a clumpy pattern with several clumps whose central regions are dominated by a massive object, the use of the mean as representative of the central value of the distribution may lead us to a wrong result because the long-edges connecting the different clumps would produce a larger edge-length average, hiding any sort of mass segregation which could be present in the data. In order to avoid this problem, we use, instead, the edge-length median, as it is more robust against the presence of outliers as the long-edges connecting well separated blobs. This strategy was formerly proposed by \cite{M&C2011}, who reformulated the procedure introducing the median instead the mean for estimating the central value of the edge-length distribution in each bin. A more detailed description of the procedure and its application can be seen in \cite{A&G2016}.

\par Summarising, we analyse the spatial pattern of the density peak using the methodology formerly proposed by \cite{allison:2010aa} but reformulated by \cite{M&C2011} to account for a possible underlying fractal pattern. This approach has been successfully applied to different observed and simulated star-forming regions \citep[i.e.][]{A&G2016,Dorvaletal2016, P&W2016, PD&E2015} and appears to be specially suitable for the analysis we plan to do with the Pipe core population.  Notice that we can apply the method to any variable, not only to mass. The two physical variables we analysed in this study are $M$ and $\rho$, both estimated for each single core in our sample. 

\section{Results and Discussion}\label{results}

\begin{figure}
\includegraphics[width=0.42\textwidth, angle=0]{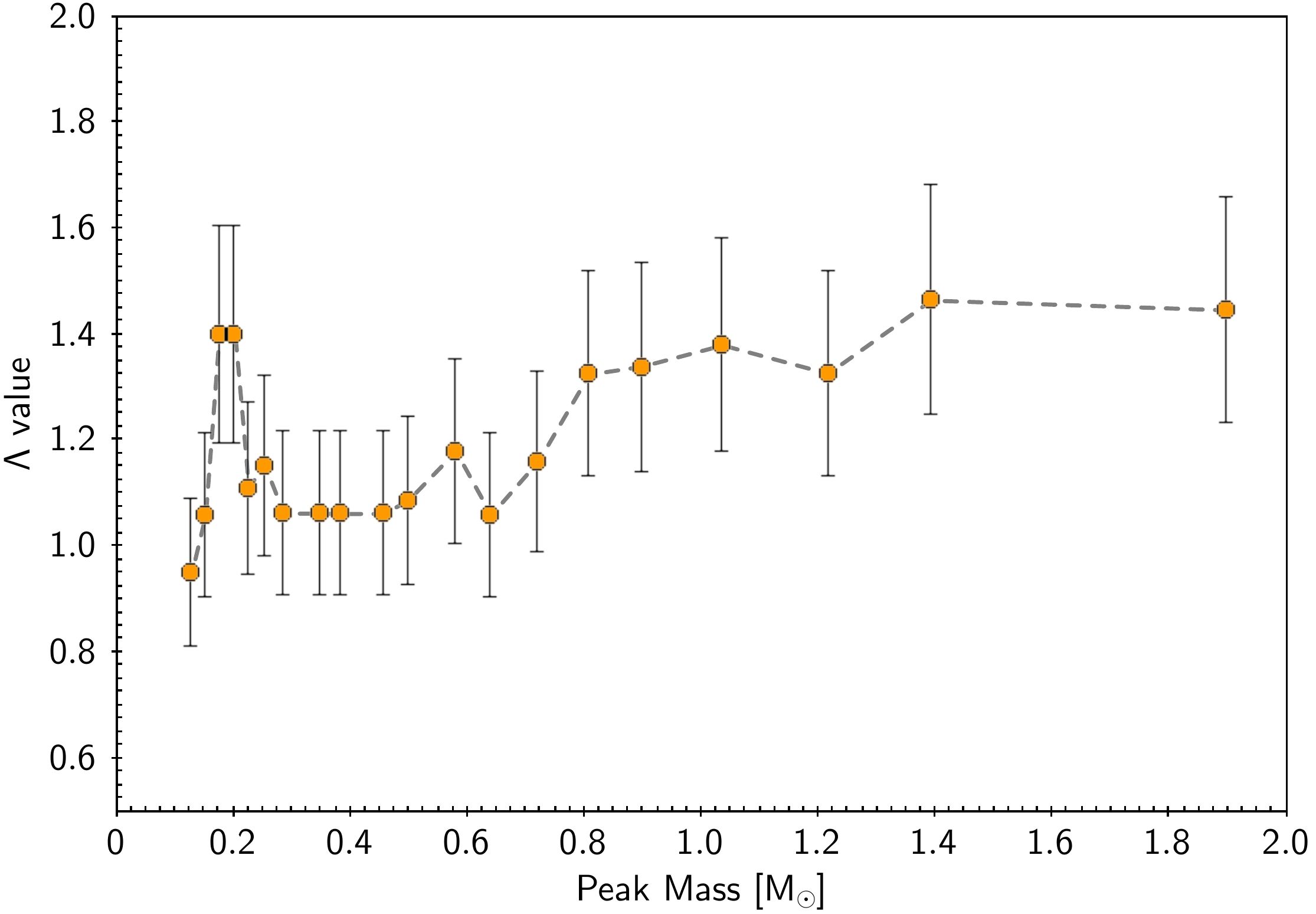}\\
\includegraphics[width=0.42\textwidth, angle=0]{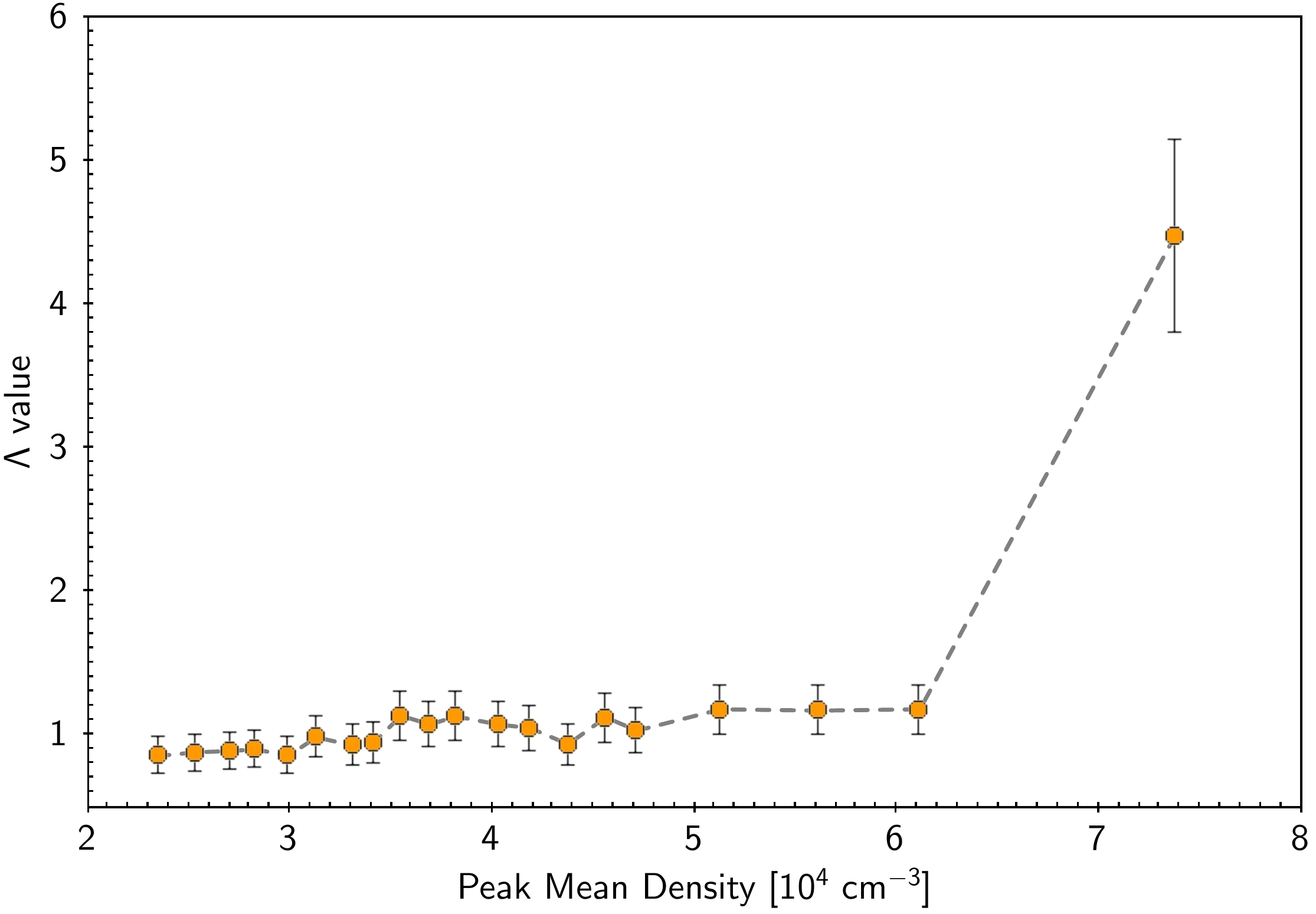}
\caption{Lambda parameter as a function of $M$ and $\rho$ for the Pipe Nebula density peak sample. We used bins of 40 elements with a sequence parameter $s$ of 10 elements. Both $M$ and $\rho$ appear to be spatially segregated but the high $\Lambda_{\rho}$, for the 40 densest cores, strongly suggests that this variable is the main responsible for drawing the geometry of Pipe's cores.}
\label{fig:lambda_40_10}
\end{figure}

\begin{figure}
\includegraphics[width=0.42\textwidth, angle=0]{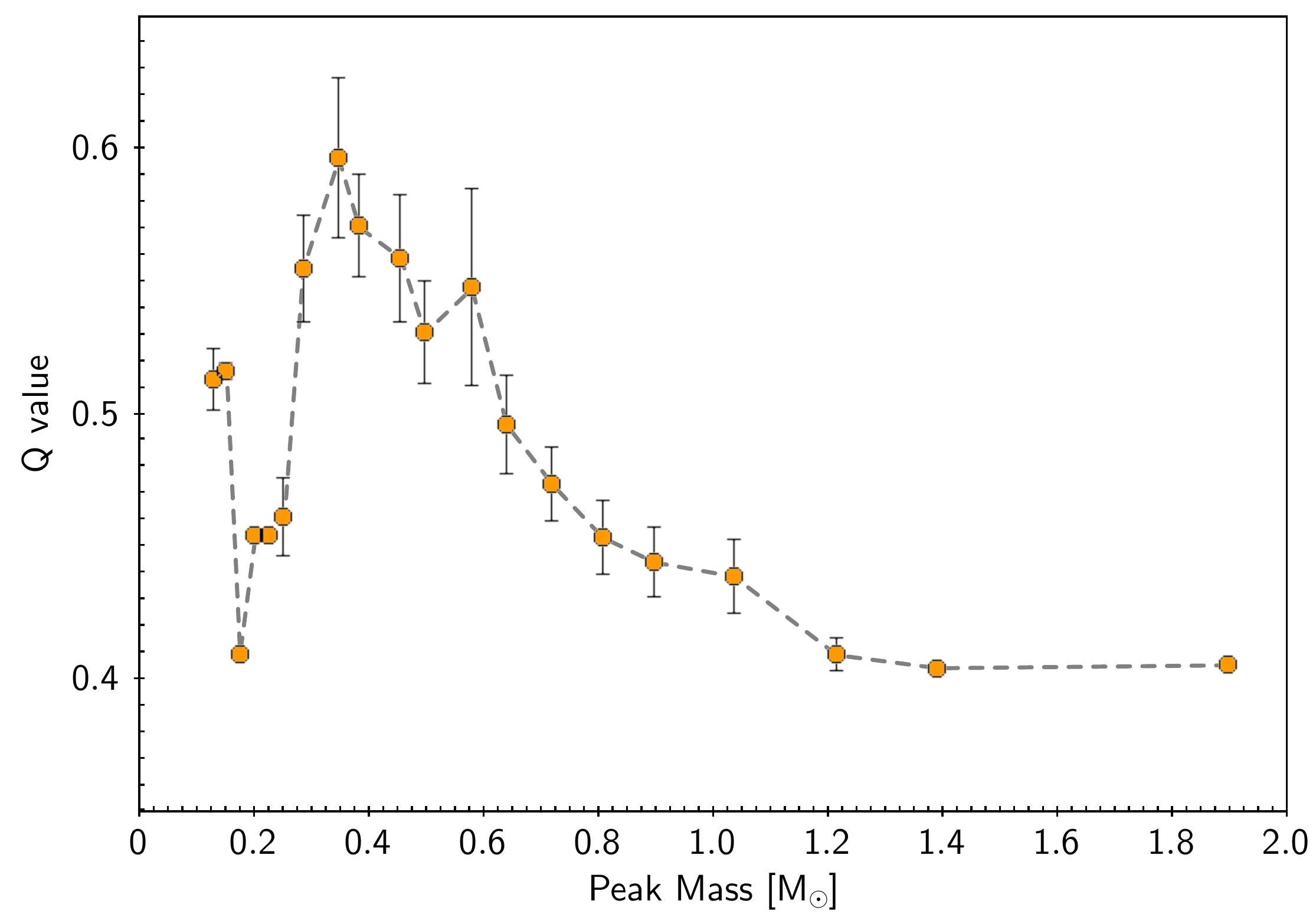}\\
\includegraphics[width=0.42\textwidth, angle=0]{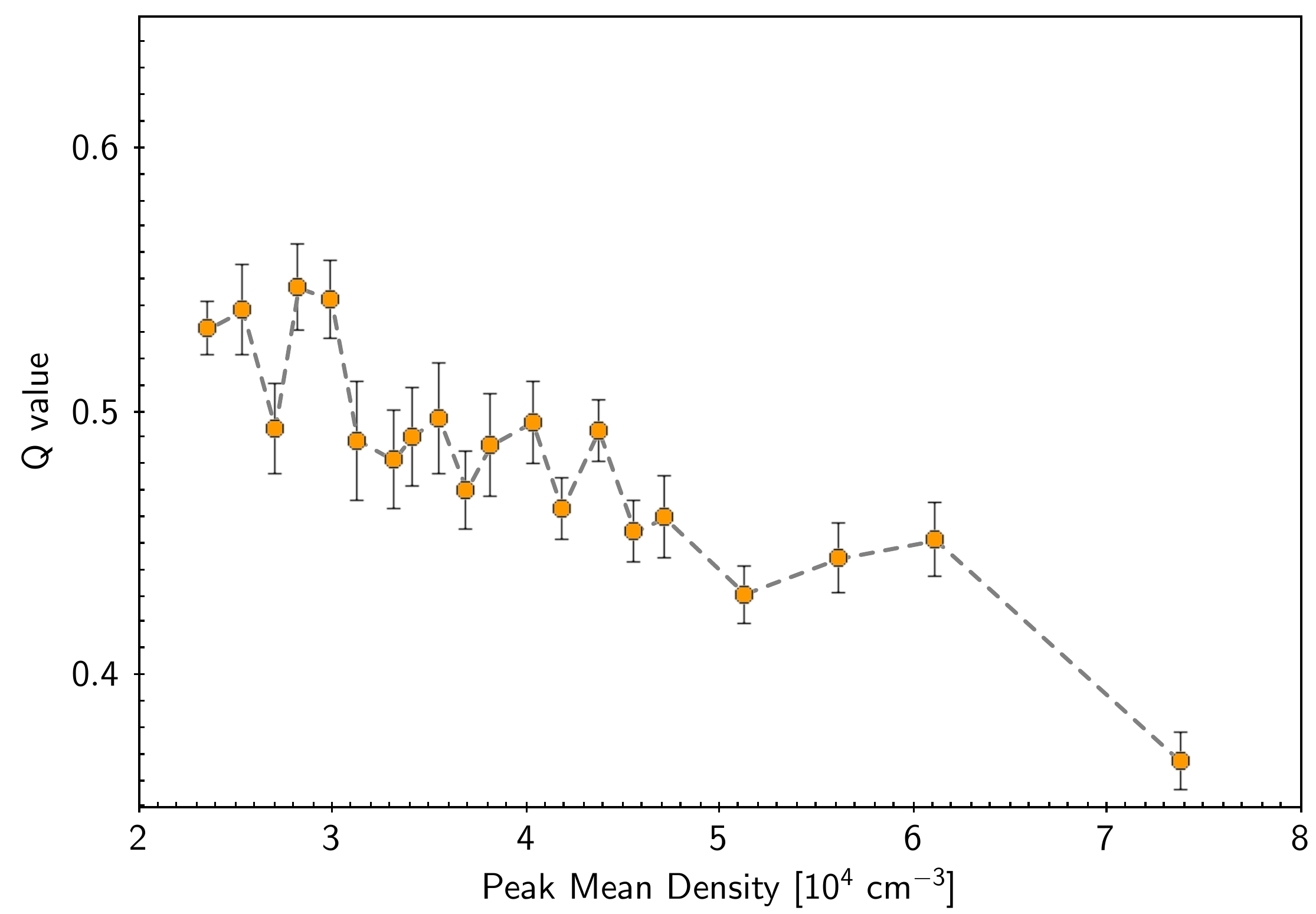}
\caption{Q-spatial parameter as a function of $M$ and $\rho$ for the Pipe Nebula density peak sample. We used bins of 40 elements with a sequence parameter $s$ of 10 elements. Error bars are estimated with jacknife resampling. $\rho$ is highly hierarchized, with Q values monotonically decreasing with increasing $\rho$.}
\label{fig:Q_40_10}
\end{figure}

\par We have applied the methodology described in section 3 to both physical variables, $M$ and $\rho$, for the column density peaks of the Pipe Nebula. A value of $n =$ 40 has been taken as the initial objects' number for the analysis of our data set, with a step of $s =$ 10. Figure \ref{fig:lambda_40_10} shows the results of this analysis for both variables. We can see that $\Lambda$ is greater than 1 (at a 2$\sigma$ confidence level) for both, the most massive and the most dense objects, but the latter are clearly more segregated than the most massive ones.

\par
\begin{figure}
\includegraphics[width=0.42\textwidth, angle=0]{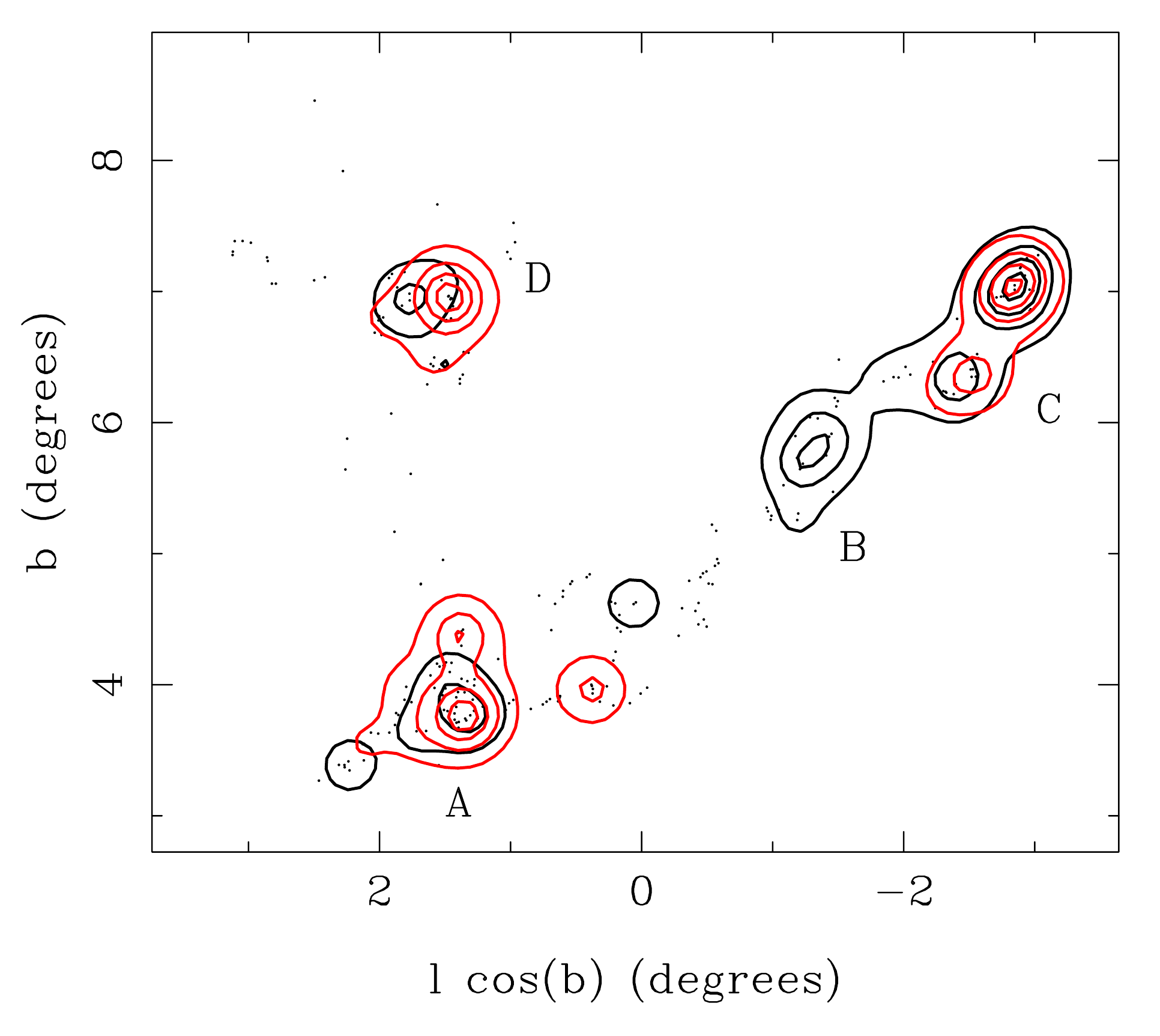}
\caption{Surface density map of Pipe nebula peaks. The contours in black show the distribution of the 40 more massive cores, while the red contours show the location of the 40 densest objects.}
\label{fig:segreg_map}
\end{figure}

 Fig. \ref{fig:Q_40_10} displays the spatial parameter $Q$, as a function of $M$ and $\rho$, using the same binning parameters. For $M$, $Q$ reaches a peak value near 0.6, but the rest of the bins show values well below of 0.8, while for $\rho$, $Q$ the peak value is near 0.55, also with most bins showing values well below 0.8. Although all $M$ intervals show a clumpy structure, the two most massive ones present the lowest $Q$ values, supporting --in the sense given by \cite{parker:2018}-- a degree of mass segregation, consistent with the $\Lambda$ analysis. Moreover, the spatial pattern shown by $\rho$ is much more clumpy, with a monotonic decreasing function of $Q$ versus $\rho$. These results are indicative of the  high level of substructure, uniformly distributed across the filaments of the Pipe, induced by $\rho$. This is consistent with the two point correlation function analysis shown by \cite{roman:2010aa}. Our quantitative analysis translates into the observable spatial plane the different distributions of either the 40 most massive cores, or the 40 densest nuclei, as shown in Figure \ref{fig:segreg_map}. Red contours correspond to the distribution by $\rho$ and black contours correspond to the one by $M$. We focus here on two main features: a) the 40 most dense cores appears to be mainly concentrated in the three vertices drawn by the filamentary distribution of Pipe's cores, and b) the 40 most massive nuclei show a not so clumpy structure, but one that is distributed in a more extended way, along main Pipe's filaments.

\par In other words, both $M$ and $\rho$ draw a hierarchical spatial distribution, where the densest and most massive absorption peaks are not distributed over the entire surface of the cloud but form singular concentrations, as seen in Figure \ref{fig:segreg_map}. Both $\Lambda$ and $Q$ plots and  the cores surface-distribution map indicate that the highest degree of spatial substructure is driven by $\rho$ and, in a lesser extension, by $M$. The former appears to be the main agent driving the primordial spatial segregation of the pre-stellar cores, at least in the Pipe Nebula. 

\par Our results support that clustering may have a primordial origin (as suggested by RAL0910), and that star cluster properties like $M$ or $\rho$ segregation, and fractal patterns may also be the result of a primordial organisation of the gas. Recent results in the Orion B Cloud \cite{parker:2018}, also using the MST method, are compatible with this vision. One has to take into account that observable substructure in young star clusters is much more complex \citep{schmeja:2008aa,sanchez:2009aa}, and processes like hierarchical merging, collaborative accretion \citep{elmegreen:2014aa}, and stellar feedback processes may play crucial roles in the observed structure of stellar clusters even at early phases of evolution \citep{ladalada:2003}.

\par Given the evolutionary state of the Pipe Nebula, very early in the formation process of a stellar system,  we can conclude that primordial $M$ and $\rho$ segregation are already present since the very initial collapse, but whereas $M$ shows a marginal segregation, $\rho$ is controlling and drawing the clumpy spatial pattern of the cores at this evolutionary time. Thus, further numerical and observational follow-ups should be also devoted to how not just the mass, but also the volume-density distribution, evolves.

\section*{Acknowledgements}

We thank to B. Elmegreen by clarifying discussions. This work has been partially funded by Spanish MINECO project AYA2016-75931-C2-1-P. C.R-Z acknowledges support from research program UNAM-DGAPA-PAPIIT IN108117, Mexico. This work is based on observations collected at the Centro Astron\'omico Hispano Alem\'an (CAHA) at Calar Alto, operated jointly by the Max-Planck Institut f\"ur Astronomie and the Instituto de Astrof\'isica de Andaluc\'ia (CSIC). This work is based in part on observations made with ESO Telescopes at the La Silla and Paranal Observatories under programs 069.C-0426 and 071.C-0324. This publication makes use of data products from the Two Micron All Sky Survey, which is a joint project of the University of Massachusetts and the Infrared Processing and Analysis Center/California Institute of Technology, funded by the National Aeronautics and Space Administration and the National Science Foundation.












\bsp	
\label{lastpage}
\end{document}